\documentclass[twocolumn,prl,amsmath,amssymb,showpacs,superscriptaddress,floatfix]{revtex4}
\usepackage{graphicx}
\usepackage{bm}
\usepackage{lineno,hyperref}
\usepackage{epstopdf}
\usepackage{epsf}
\begin{document}
\title{Anomalous behavior of dispersion curves in water-like systems and water}

\author{Yu. D. Fomin, E. N. Tsiok, V. N. Ryzhov, and V. V. Brazhkin}
\affiliation{ Institute for High Pressure Physics RAS, 108840
Kaluzhskoe shosse, 14, Troitsk, Moscow, Russia}

\date{\today}

\begin{abstract}
In the present paper we consider dispersion curves of longitudinal
excitations of a model core-softened liquid and SPC/E model of
water. We show that both systems demonstrate anomalous behavior of
the excitation frequencies: the frequencies of the excitations can
decrease with temperature along isochors, while in normal liquids
they should increase. This observation allows to introduce one
more water anomaly - anomalous dependence of excitation
frequencies on temperature.
\end{abstract}


\maketitle

\section{Introduction}

It is widely known that many properties of crystals can be
efficiently described in frames of their collective excitations -
phonons \cite{solidstate,kittel}. The phonon approach allows to
predict many important properties of crystalline solids such as,
for instance, the stability of a crystal or its heat capacity. In
this respect it is also important to mention that investigation of
collective excitations makes a link between dynamic and
thermodynamic properties of matter.

The situation becomes more complex in the case of liquids.
Although the collective excitations in liquids were extensively
studied both theoretically \cite{boon} and experimentally
\cite{ruocco}, the absence of a simple exactly solvable model like
the model of a harmonic crystal, prevented the appearance of a
theory of liquid based on the collective excitations. Only few
years ago some publications appeared which tried to explain the
behavior of the heat capacity of liquids basing on collective
excitations \cite{phonon-liq}. Although the results presented in
this paper are very encouraging, this topic definitely requires
further development.

Two topics on collective excitations in liquids got most of
attention. First of all, it is the existence of transverse
excitations in liquids. It was believed for a long time that
liquids do not support the transverse excitations. At the same
time it was widely known that liquids with very high viscosity do
demonstrate transverse waves. Moreover, later it was recognized
that even the liquids with low viscosity can sustain transverse
excitations. For example, in Ref. \cite{lj4} dynamical properties
of Lennard-Jones model were discussed. The results of this paper
predicted the existence of transverse excitations in the system.
Nowadays the transverse excitations were observed experimentally
in liquids metals
\cite{hos-fe-cu-zn,psd-ga,psd-fe,psd-sn,psd-fe-1,psd-sn-1} and
some other liquids. So the existence of transverse waves  can be
considered as strictly proved.

Another point of interest is the so called positive sound
dispersion (PSD). The PSD means that the frequency of longitudinal
excitations exceeds the frequency obtained from the Debye law
$\omega_D(k)=c_s \cdot k$, where $c_s$ is the adiabatic speed of
sound. PSD also was observed experimentally in many different
systems, for instance, liquid metals
\cite{hos-fe-cu-zn,psd-ga,psd-fe,psd-sn,psd-fe-1,psd-sn-1}.
Importantly, both PSD and transverse excitations disappear when
the temperature is risen or the density is decreased. It allowed
to propose that these properties can be used to demarcate
liquid-like and gas-like regimes in fluids \cite{frpre,frprl,ufn}.

Even further complication can be expected in the case of so called
"anomalous liquids", i.e. the liquids which demonstrate unusual
properties. The most well known anomalous liquid is water which
demonstrates more then 70 anomalies, for instance, density
anomaly, diffusion anomaly, maxima of thermodynamic response
functions, etc (one can find an extensive description of most of
the anomalous properties of water at the site \cite{wateranom}).
One can see that water demonstrates both thermodynamic and dynamic
anomalies. However, up to now we are not aware of any systematic
research of possible anomalous behavior of collective excitations
in water.

Although the behavior of water is of extreme interest it appears
to be very complex. Up to now there is no any model of water which
can correctly describe many different properties (for example,
phase diagram, viscosity, the location of density maxima, etc.)
simultaneously. In Refs. \cite{vega-comparison,vega-comparison1} a
critical comparison of several common water models is given. The
authors give the highest scores to the TIP4P/2005 model
\cite{tip4p2005}, however, even this model strongly underestimates
the melting temperatures at different pressures
\cite{vega-comparison}.

For the reason of the difficulties described above it is often
reasonable to start investigations from simple models which can
demonstrate some properties similar to water, for example, density
or diffusion anomaly. Investigation of such systems can give a
deep insight into the nature of some phenomena observed in water
and other anomalous liquids being computationally cheaper then
studies of the water itself. In Ref. \cite{stanley} physical origin of
polyamorphism of ice is discussed. A simple model  able to reproduce the
coexistence of two disordered phases was proposed. In this model two tetrahedra
of water (4 molecules complexes) interact with each other via a core-softened potential.
Although core-softened systems fail to reproduce the behavior of water exactly they
are able to mimic many unusual water properties, such as density anomaly, diffusion
anomaly, structural anomaly, maxima of the response functions, etc.

Core-softened systems received a lot attention in the literature. Many different systems were
proposed and studied. It was found that these systems demonstrate numerous unusual features:
complex phase diagram, maxima on the melting line, unusual phases (for example, diamond phase
in a system with isotropic pair potential \cite{star,s135a}), water-like anomalies, etc. Moreover, the behavior
of the system is very sensitive to the shape of the interaction potential which allows to
modify the system in order to reproduce a feature of interest. For example, in Ref. \cite{erdeb}  it was
shown that the regions of density, diffusivity and structural anomalies in water form nested domains in
the density-temperature plane. In Ref. \cite{sio2} the same work was performed for liquid silica and it was discovered that the
mutual location of the anomalies is different from the case of water. In Refs. \cite{silica-anom-1,silica-anom-2} a system with a core-softened
potential was studied. The functional form of the potential was remained constant, but its parameters
varied. It was shown that in frames of the same model one can obtain the location of the anomalous
regions like in water or like in silica. Similar effects of changing of some anomalous regions are
observed when comparing different tetrahedral liquids, for example, water, silicon and germanium modeled by
Stillinger-Weber potentials \cite{ind1}.


A model which demonstrates complex behavior with numerous
anomalies was introduced in our previous works \cite{s135a,s135}.
Below we will call it Repulsive Shoulder System (RSS) model. This is a
system of particles interacting via a potential is given by the formula:

\begin{equation}\label{pot}
U(r)/ \varepsilon = \left( \frac{\sigma}{r}
\right)^{14}+0.5\left(1- tanh(k(r-\sigma_1)) \right),
\end{equation}
with $k=10.0$ and the parameter $\sigma_1$ determines the width of
the repulsive shoulder. It is convenient to express all quantities
in the units of the potential, i.e. the parameter $\varepsilon$ is used
as a unit of energy and $\sigma$ as a unit of length. All other quantities
can be expressed from these parameter. Below all quantities are given in these
reduced units. In our previous publications it was shown
that this system demonstrates very complex behavior
\cite{s135,s135a,s135b,s135c,s135d,s135e}. Moreover, the behavior
of the system is extremely sensitive to the magnitude of the
parameter $\sigma_1$. In particular, if $\sigma_1=1.35$ then the
system demonstrates the diffusion anomaly, the density anomaly and
the structural anomaly. In the case of $\sigma_1=1.55$ the
diffusion anomaly disappears and with $\sigma_1=1.8$ the density
anomaly also disappears \cite{s135d}.

In the present paper we investigate the dispersion curves of
longitudinal excitations of the RSS system and show that they
demonstrate anomalous behavior: while in normal liquid the
excitation frequency increases upon isochoric heating, in RSS the
excitation frequency can decrease with isochoric temperature
increase. We demonstrate that this anomalous behavior cannot be
reproduced in the simplest theoretical approximation -
Hubburd-Beeby model which is widely used for description of the
collective excitations in liquid metals. Finally, we perform
preliminary study of dispersion curves of water and show that they
also demonstrate the same anomaly.

\section{System and Methods}

We simulated a system of 4000 particles in a cubic box with
periodic boundary conditions interacting via the potential
(\ref{pot}). We performed simulations along three distinct
isochores: $\rho=0.5$, $\rho=0.65$ and $\rho=1.15$. The first
isochore crosses the region of the water-like anomalies of the
system. The second isochore is at slightly higher density then the
anomalies reported earlier. Finally, the $\rho=1.15$ isochore
belongs to the high-density regime where no anomalies are
expected. The temperatures varies from $T_{min}=0.1$ up to
$T_{max}=1.0$ at $\rho=0.5$, from $T_{min}=0.3$ up to
$T_{max}=1.5$ at $\rho=0.65$ and from $T_{min}=2.0$ up to
$T_{max}=10.0$ at $\rho=1.15$.

The system was simulated by means of molecular dynamics method.
Firstly the system was equilibrated in NVT (constant number of
particles N, volume V and temperature T) ensemble for $10 \cdot
10^6$ steps. The time step was set to $dt=1 \cdot 10^{-4}$. After
the equilibration the system was propagated in microcanonical NVE
(constant number of particles N, volume V and internal energy E)
ensemble. In case of $\rho=0.5$ the propagation period is $2 \cdot
10^8$ steps. At other densities it is $1 \cdot 10^8$ steps. In
order to improve the statistics we divide the propagation periods
into the blocks of $5 \cdot 10^6$ steps and perform block
averaging \cite{book_fs}.

In order to find out the dispersion curves we calculated the
longitudinal part of the velocity flux correlation function:

\begin{equation}
C_L(k,t)=\frac{k^2}{N}\langle J_z({\bf k},t) \cdot J_z(-{\bf
k},0)\rangle
\end{equation}
where $J({\bf k},t)=\sum_{j=1}^N {\bf v}_j e^{-i{\bf k r}_j(t)}$
is the velocity current and the wave vector $\bf{k}$ is directed
along the z axis \cite{hansenmcd,rap}. Note that the wave vector $\bf{k}$ is
not related to the parameter $k$ of the interaction potential. We believe that
this coincidence in notation does not lead to ambiguity. The excitation frequencies
$\omega$ were determined as peaks of the Fourier transform of
$C_L$.

In spite of large simulation time and rather good averaging the
raw spectra are still noisy. In order to remove this noise we
approximate the spectra by the following equation:

\begin{equation}\label{fit}
 \omega(k)= c_s k+ b_1 k^{5/2}+b_2k^{11/4}+b_3k^{23/8},
\end{equation}
where $c_s$ is adiabatic speed of sound and $b_1$, $b_2$ and $b_3$
are fitting coefficients \cite{sp-mct}. This formula is obtained
in frames of Mode-Coupling Theory (MCT) treatment of collective
excitations in fluids. The adiabatic speed of sound was calculated
as $c_s=\left(\gamma \left( \frac{\partial P}{\partial \rho}
\right)_T \right)^{1/2}$, $\gamma= c_P/c_V$,
$c_P=c_V+\frac{T}{\rho ^2} \left( \frac{\partial P}{\partial T}
\right)_{\rho}^2 \left( \frac{\partial P}{\partial \rho}
\right)_T^{-1}$, $c_P$ and $c_V$ are isobaric and isochoric heat
capacities.

We also perform simulation of SPC/E model of water. We study the
system of 4000 water molecules in a cubic box under PBC. The
system is equilibrated for 0.5 ns in NVT ensemble. The time step
is set to $0.1$ fs. 2 blocks of $3 \cdot 10^6$ steps are made for
calculating of averages. NVE ensemble is used at this stage The
calculation of the dispersion curves is made in the same fashion
as in the case of RSS system.

We also give a plot of the dispersion of soft sphere system, i.e.
the system of particles interacting via power law potential
$U_{soft}=\varepsilon \left( \frac{\sigma}{r} \right)^n$. This
system is a generic model of a simple liquid and it does not
demonstrate any kind of anomalies. The simulation setup of this
system is given in our previous publication \cite{frprl}.

Lammps simulation package is used for all simulations of this
paper \cite{lammps}.

\section{Results and Discussion}

\begin{figure}
\includegraphics[width=8cm]{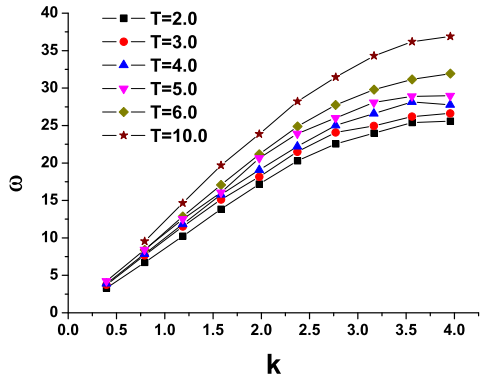}%

\caption{\label{softsp}Dispersion curves of the soft sphere system with $n=12$
at $\rho=1.0$ isochore.}
\end{figure}

In the beginning of the discussion we give example of the
dispersion curves of soft sphere system with $n=12$ at $\rho=1.0$
and a set of temperatures (Fig. \ref{softsp}). Soft spheres is a system which
can serve as etalon os a simple liquids, and no anomalies is expected in soft spheres.
One can see that in a simple liquid the frequencies of excitations increase with
temperature. Below we call this type of dependence as normal while
decrease of the frequencies with temperature as anomaly.

Let us consider the dispersion curves of RSS at the density
$\rho=0.5$ and different temperatures. As it is shown in the
supplementary materials the dispersion curves are well fitted by
the eq. \ref{fit}. Therefore, here we show only the fitted curves.



\begin{figure}
\includegraphics[width=8cm]{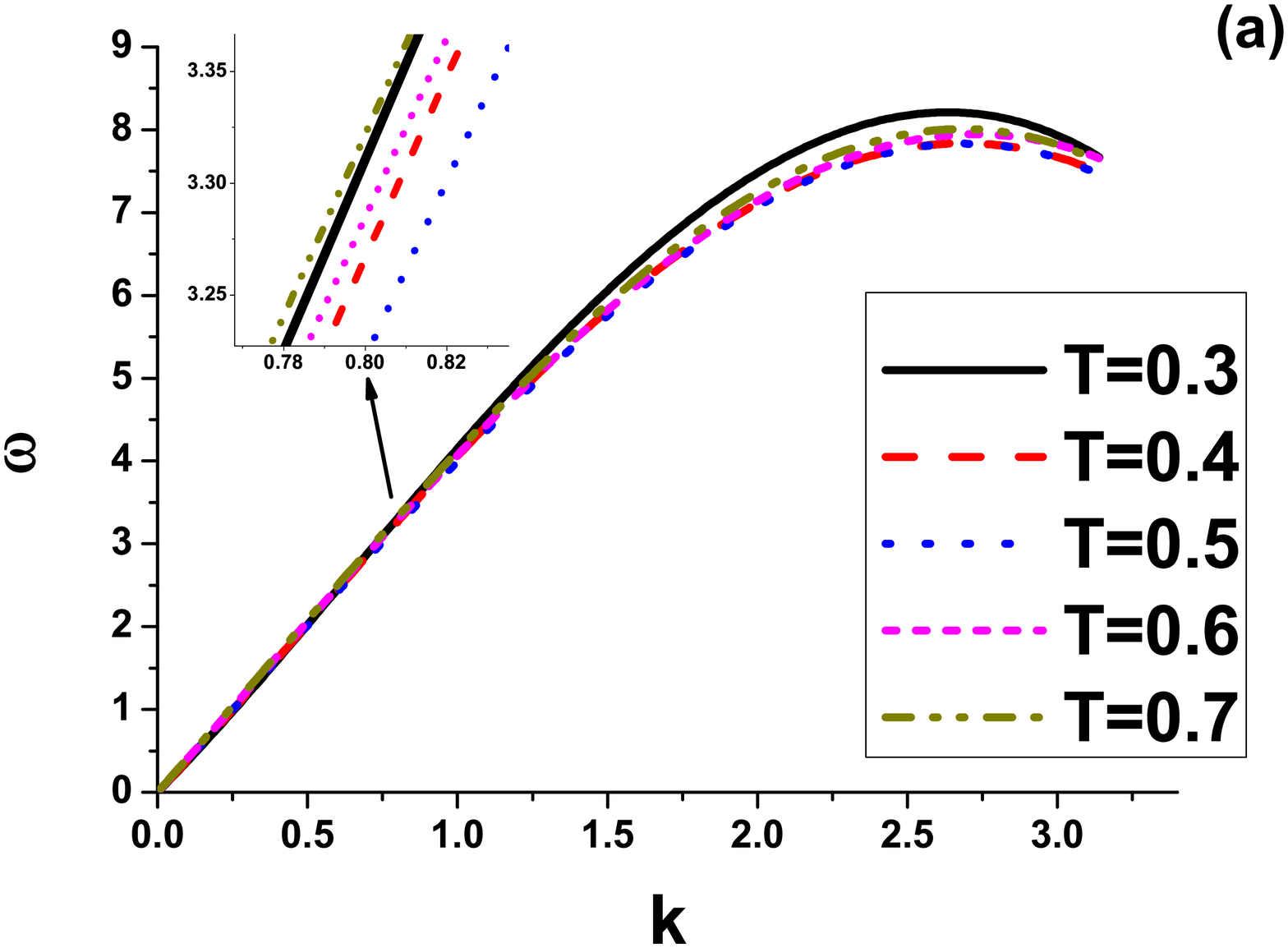}%

\includegraphics[width=8cm]{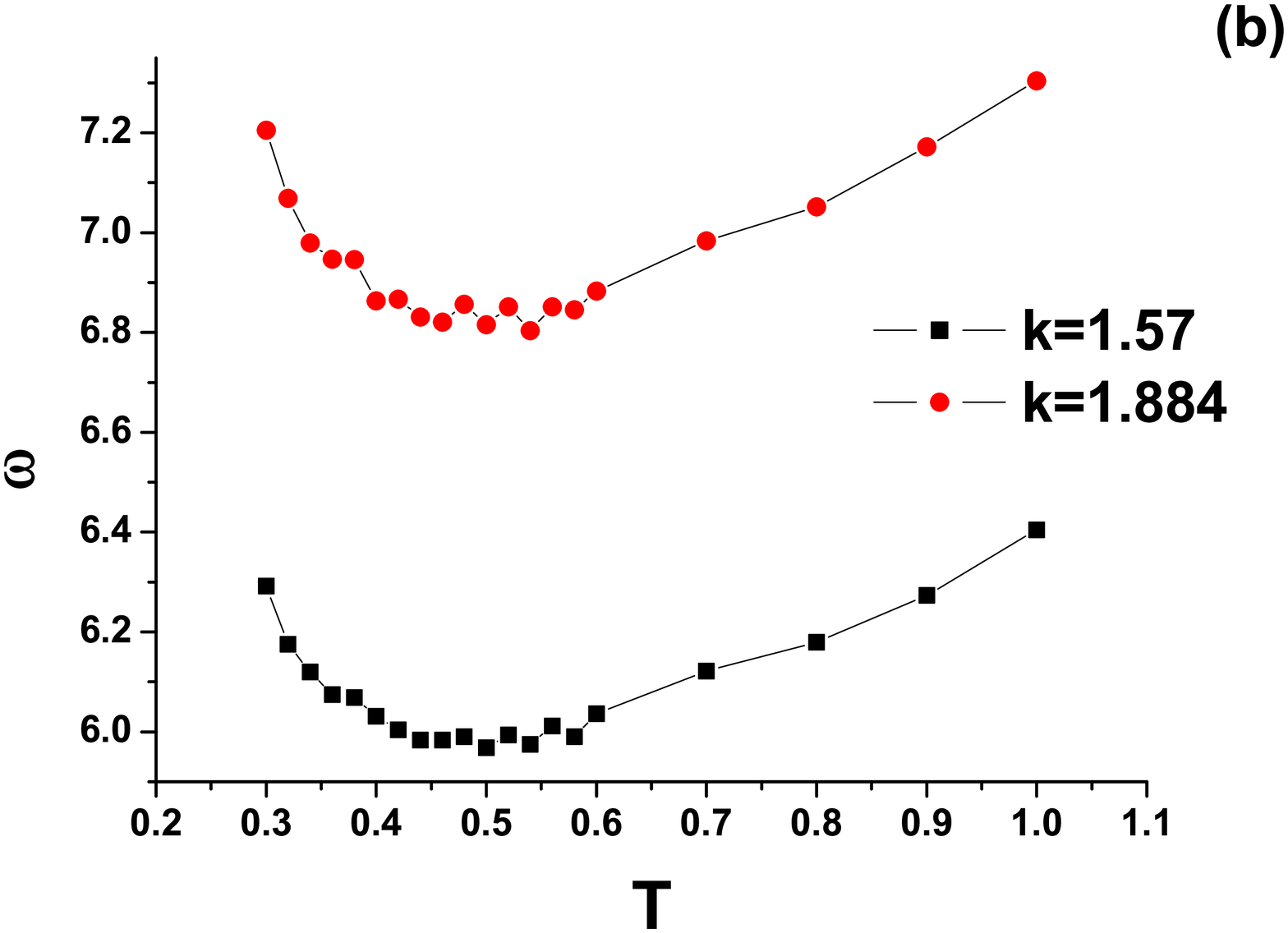}%

\caption{\label{r05-anom}(a) Dispersion curves of the system at $\rho=0.5$ which
demonstrate the anomalous decrease of the frequency with
temperature. The inset enlarges the region of $k$ around $0.8$.
(b) Temperature dependence of the frequency at two magnitudes of
the wave-vector $k$.}
\end{figure}

Fig. \ref{r05-anom} (a) shows the dispersion curves of the system
at $\rho=0.5$ and a set of temperatures. One can clearly see that,
for instance, the curve at $T=0.4$ is below the one at $T=0.3$. It
means that the frequency decreases with temperature. However, if
the temperature is increased further the frequencies start to
increase with temperature as in a normal liquid. This behavior is
qualitatively different from the one of soft spheres shown above.
Fig. \ref{r05-anom} (b) demonstrates the temperature dependence of the excitation
frequency at fixed magnitude of the wave vector. Although the data
are still a bit noisy one can clearly identify a minimum at the
temperature around $T=0.5$. Therefore, one can say that the
excitation spectra of the system are anomalous if from $T=0.3$ (the lowest temperature we studied) up to $T=0.5$ and
behave like in normal liquid above this temperature.

We believe that anomalous dependence of the dispersion curves on
temperature requires also theoretical interpretation. The simplest
approach to the excitation spectra of liquids is the Hubbard-Beeby
(HB) model \cite{hbmodel}. However, it does not reproduce the
anomalous temperature dependence of the frequencies (see the
supplementary materials for the discussion). More elaborate
theories are required.

We remind that at $\rho=0.5$ the system demonstrates the density
anomaly \cite{s135,s135a,s135b,s135c,s135d,s135e} which leads to
appearance of a minimum on the pressure dependence on temperature
at this isochore. However, the adiabatic speed of sound $c_s$
along the same isochore is a monotonous function of temperature
(see the supplementary materials for the plots of the $P$ and
$c_s$ along the isochore). In this respect it is interesting to
consider the PSD in this system. PSD is the phenomenon when the
frequency $\omega (k)$ exceeds the Debye value $\omega_D (k)= c_s
k$. Therefore, in the present case we observe two tendencies: the
frequencies become smaller while the speed of sound increases.
However, at $T>0.5$ the frequencies change to increasing with
temperature. Therefore, one can expect some kind of non-monotonous
behavior of PSD.





\begin{figure}
\includegraphics[width=8cm]{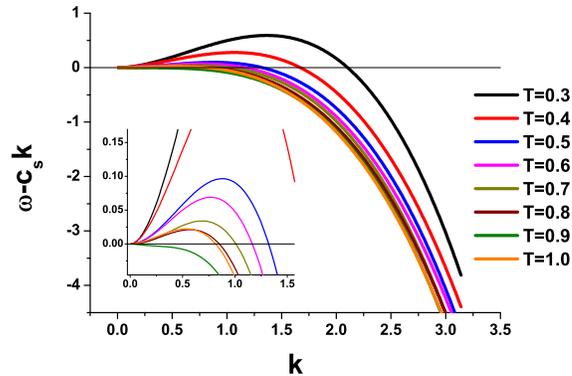}%

\caption{\label{r05-psd}PSD at $\rho=0.5$. The inset enlarges the region of small
k. The horizontal line marks the zero level.}
\end{figure}

In order to study the PSD in the system we consider the difference
$\omega(k)-c_sk$ (Fig. \ref{r05-psd}). The presence of PSD means
that this quantity has positive magnitude. One can see that the
PSD disappears somewhere between $T=0.8$ and $0.9$, but reappears
again at $T=1.0$. Therefore, the non-monotonous behavior of PSD
does appear in this system. Such disappearance of PSD was already
observed in our recent publication \cite{we-fpe}, however, no
explanation was given in that work. Now we see that the reason of
such complex behavior is the anomalous behavior of the excitation
frequencies.

\begin{figure}
\includegraphics[width=8cm]{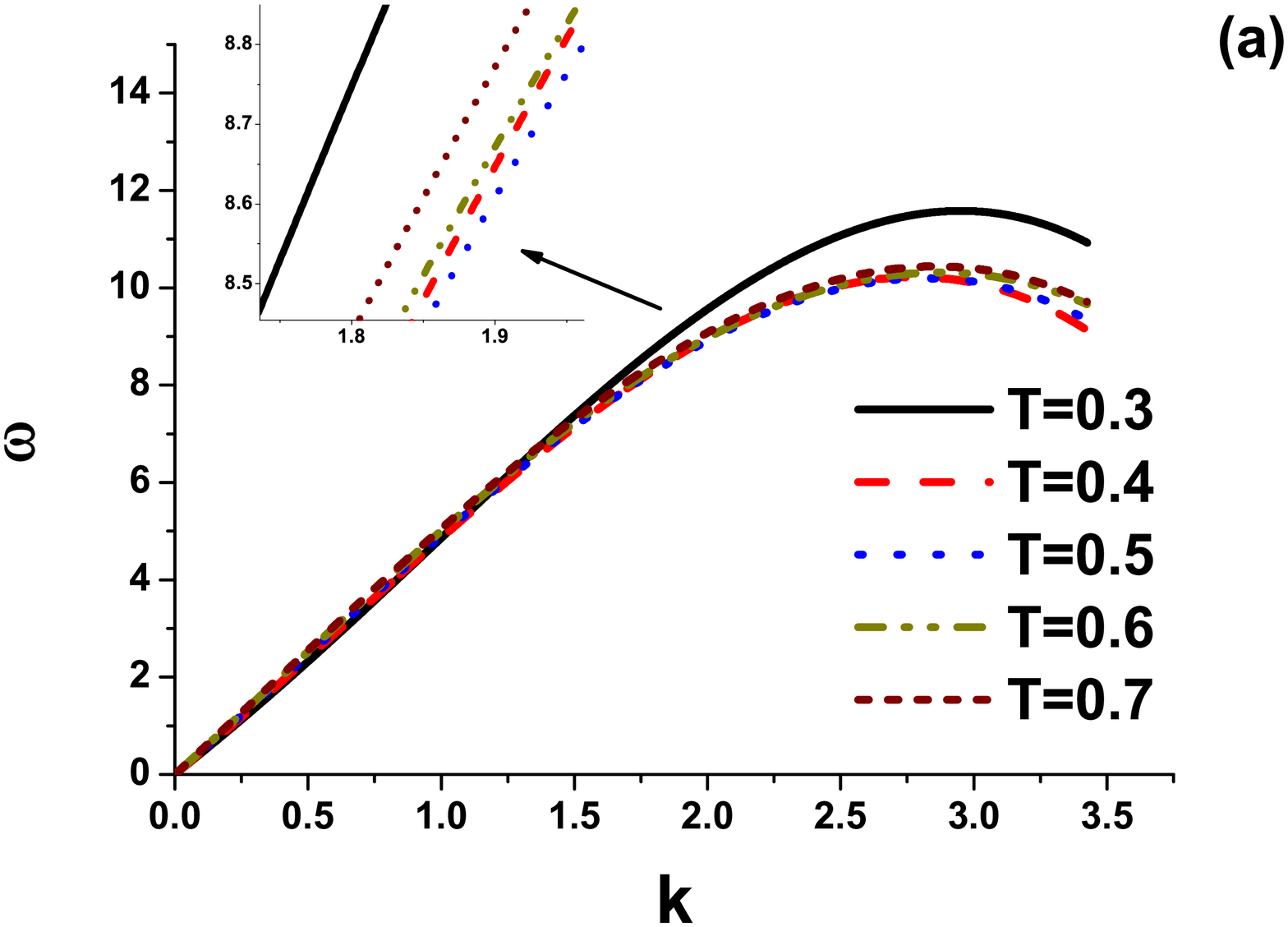}%

\includegraphics[width=8cm]{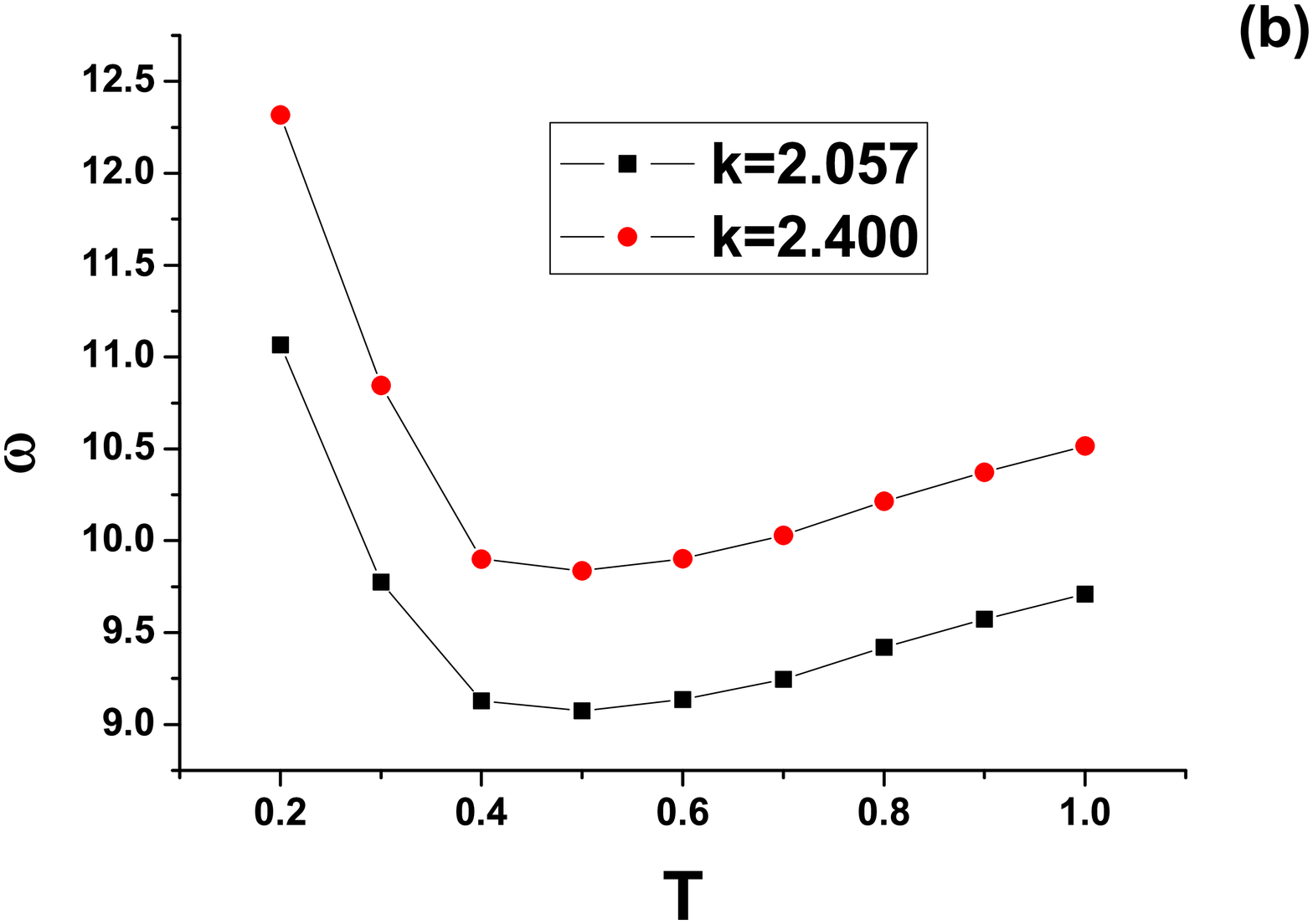}%

\includegraphics[width=8cm]{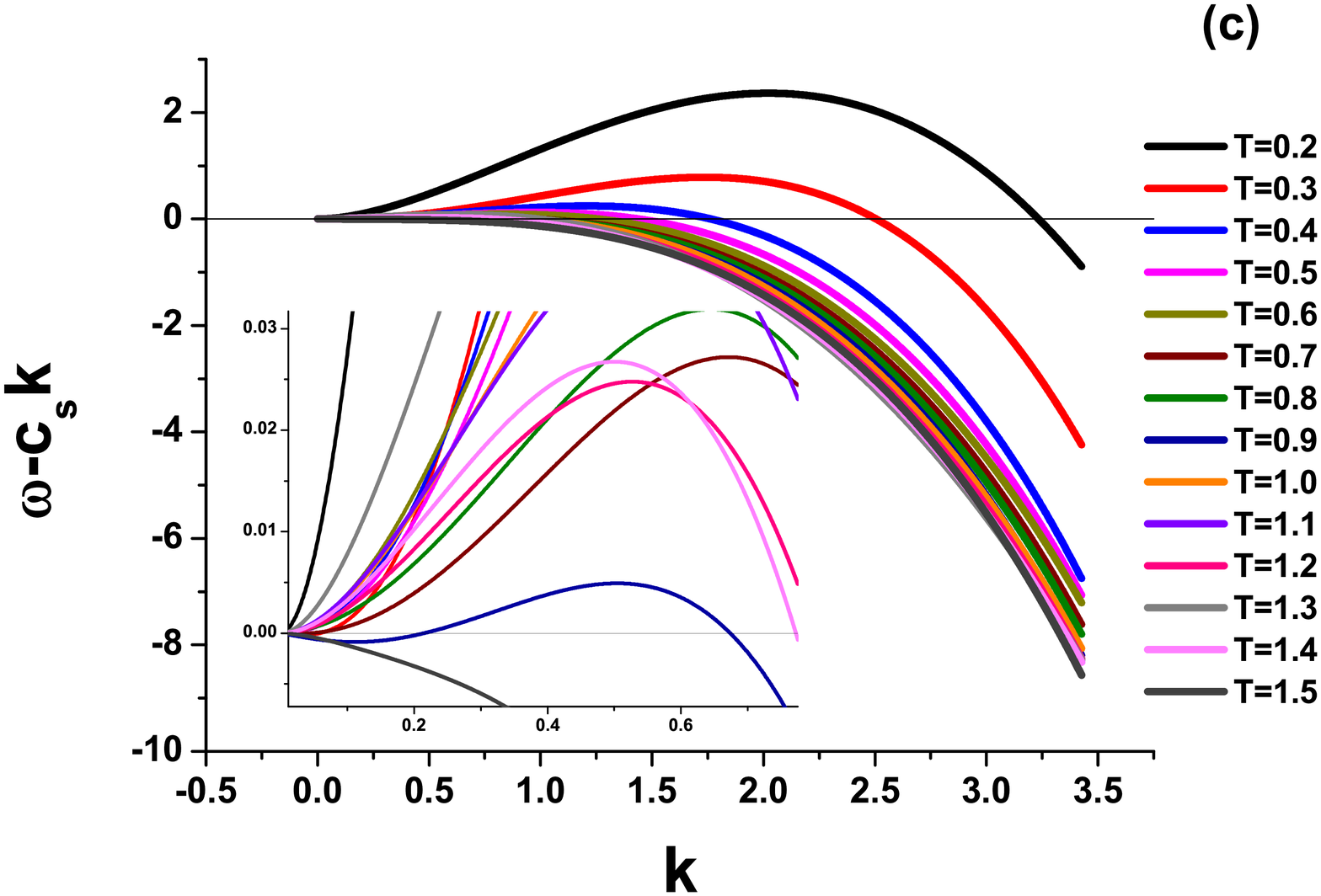}%

\caption{\label{r065}(a) Dispersion curves at density $\rho=0.65$. The inset
enlarges the region $k \approx 1.85$. (b) Temperature dependence
of excitation frequency at fixed wave-vector. (c) PSD in the
system.}
\end{figure}

The same calculations were repeated for two more densities:
$\rho=0.65$ and $\rho=1.15$. Fig. \ref{r065} shows the dispersion
curves (panel (a)), the temperature dependence of excitation
frequencies at a fixed wave vector (panel (b)) and the PSD in the
system at $\rho=0.65$. One can see that the behavior of the
dispersion curves at this density is qualitatively equivalent to
the case of $\rho=0.5$: there is an anomalous region of
temperatures where the excitation frequency decreases with
temperature. A well pronounced minimum is observed at panel (b) at
$T \approx 0.5$. At the same time the PSD has complex dependence
on temperature at temperatures from $T=1.0$  up to $T=1.5$. The
PSD becomes smaller or larger with temperature. However, such
irregular behavior of PSD can be also a result of inaccuracies of
calculations.

In the case of $\rho=1.15$ the behavior of the dispersion curves
is normal, i.e. frequencies increase with temperature. The
corresponding plots are given in the supplementary materials.


Importantly, the anomalous behavior of the system under
consideration is related to the presence of two length scales in
the potential \ref{pot}. A competition between these length scales
takes place at intermediate densities
\cite{s135,s135a,s135b,s135c,s135d,s135e}. The density $\rho=1.15$
is rather high and the behavior of the system is dominated by the
core of the potential $\sigma$. Because of this no anomalies is
expected at this density which is confirmed by the results of
simulations.



\bigskip

\begin{figure}
\includegraphics[width=8cm]{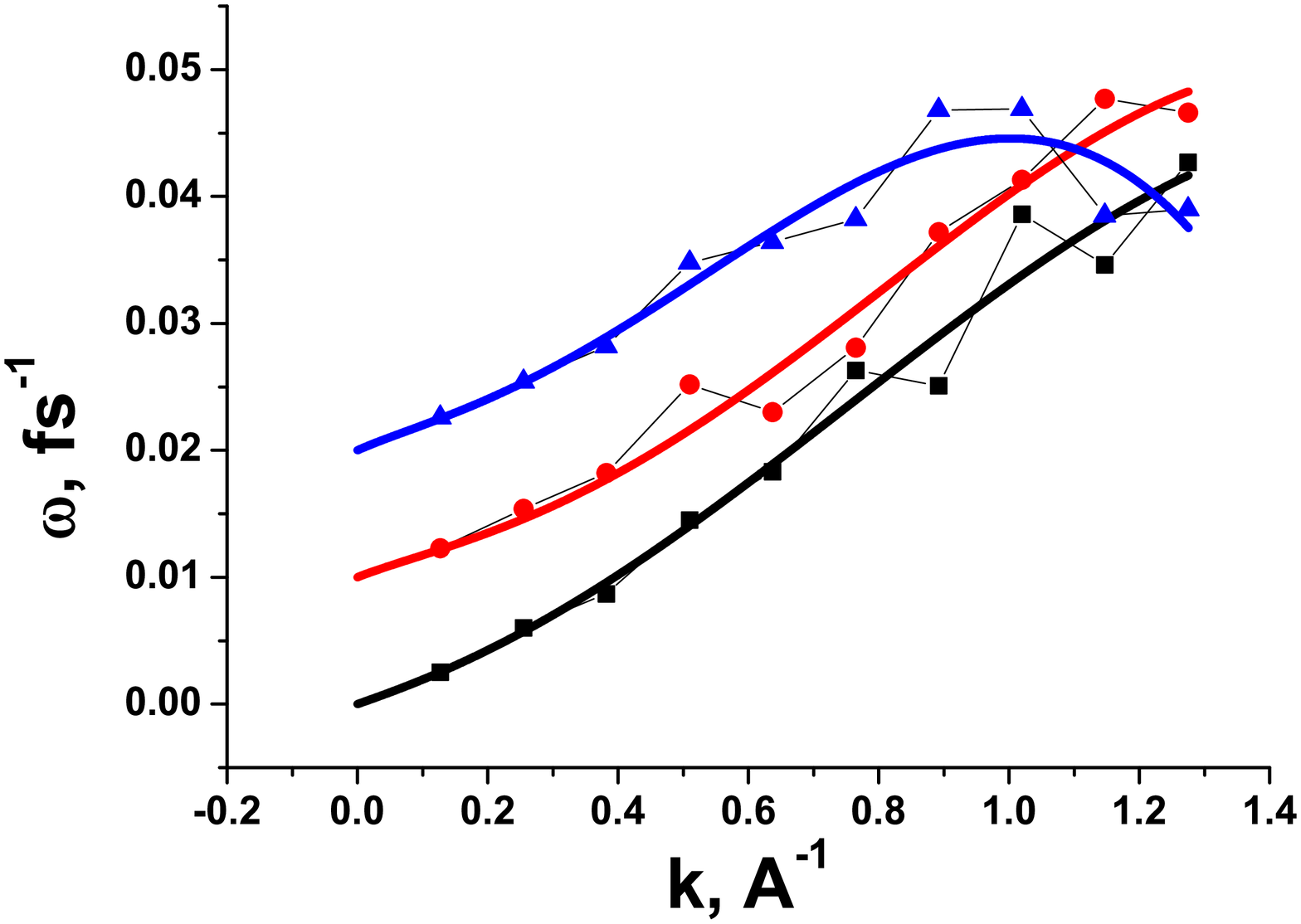}%

\caption{\label{water-app}Examples of dispersion curves of water and their
approximation by MCT formula (eq. \ref{fit}). From bottom to top:
$T=400$ K, $500$ K and $600$ K. The curves are shifted by 0.01
with respect to the previous one in order to make the plot
clearer. The symbols give the MD results and the curves the
results of fitting.}
\end{figure}

\begin{figure}
\includegraphics[width=8cm]{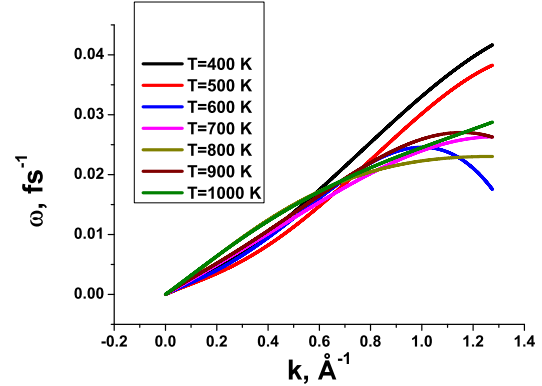}%

\caption{\label{water-all}Dispersion curves of water at $\rho=1$ $g/cm^3$ and different temperatures.}
\end{figure}

The system with potential \ref{pot} demonstrates a number of
water-like features, like, density anomaly, diffusion anomaly,
structural anomaly, maxima of response functions, etc. Basing on
this one can guess that presence of the dispersion curve anomaly
in the RSS system can be also similar to the behavior of water and
that water also demonstrates the dispersion curve anomaly. In
order to check it we performed simulations of the SPC/E model of
water. Since the simulation of water is more time-consuming we
collected much less statistics and the spectra are much noisier
comparing to the ones of RSS system. We employ the same MCT
fitting of the spectra (eq. \ref{fit}) in order to remove the
noise. Several examples of dispersion curves are shown in Fig.
\ref{water-app}. One can see that the MD results are rather noisy,
but the fitting works quite well. Fig. \ref{water-all} shows the
dispersion curves of water at a set of temperatures along $\rho=1$
$g/cm^3$ isochor. Only the results of fitting are shown. Comparing to the
case of RSS model the dispersion curves look more complex. At low temperatures
one can observe a bend of the curves downward at low k vectors which disappears
at high temperatures and the curves intersect at intermediate wave vectors.
Because of this we compare the frequencies at the highest wave vectors, which are
close to the Brillouin zone boundary. In the case of liquids the boundary of
Brillouin zone is a half of the wave vector corresponding to the first maximum of the
structure factor. One can see that at the temperatures below $600$ K the excitation
frequencies decrease with temperature. Therefore, the same type of
anomaly is observed in water too. However, this conclusion
requires more elaborate calculations, since the dispersion curves of water
are more complex then the ones of the RSS model.

As it was mentioned in Introduction, water demonstrates many
different anomalies. The results of the present paper discover one
more anomaly of water: anomalous dependence of the excitation
frequencies on temperature.

In conclusion, we consider the spectra of longitudinal excitations
in a core-softened system (RSS model) and we find that at some
densities the dispersion curves demonstrate anomalous temperature
dependence of the excitation frequencies. The same kind of anomaly
is also observed in SPC/E model of water which allows to say thay
a novel anomaly of water is discovered. Further investigations are
required in order to find out the reasons of this anomaly.

This work was carried out using computing resources of the federal
collective usage center "Complex for simulation and data
processing for mega-science facilities" at NRC "Kurchatov
Institute", http://ckp.nrcki.ru, and supercomputers at Joint
Supercomputer Center of the Russian Academy of Sciences (JSCC
RAS). The work was supported by the Russian Foundation of Basic Research (Grants No 18-02-00981).

\bigskip
\newpage

\section{Anomalous behavior of dispersion curves in water-like systems and water: supplementary materials.}

\author{Yu. D. Fomin, E. N. Tsiok, V. N. Ryzhov, and V. V. Brazhkin}
\affiliation{ Institute for High Pressure Physics RAS, 108840
Kaluzhskoe shosse, 14, Troitsk, Moscow, Russia}

\maketitle

\date{\today}

\section{Fitting of the dispersion curves}

As it is discussed in the main text, the dispersion curves are
fitted by the Mode-Coupling Theory (MCT) equation:

\begin{equation}\label{fit}
 \omega(k)= c_s k+ b_1 k^{5/2}+b_2k^{11/4}+b_3k^{23/8},
\end{equation}
where $c_s$ is adiabatic speed of sound and $b_1$, $b_2$ and $b_3$
are fitting coefficients \cite{sp-mct}.

\begin{figure}
\includegraphics[width=8cm]{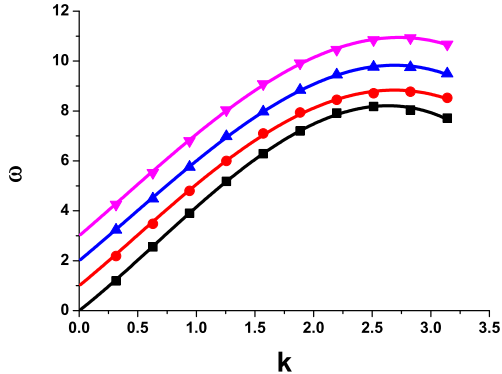}%

\caption{\label{fit-fig} Dispersion curves of the system at $\rho=0.5$ and (from
bottom to top) $T=0.3$, $0.4$, $0.5$ and $0.6$. Each upper curve
is shifted by unity from the previous one to make the figure
clearer. The symbols denote the data from simulation while the
lines are the fitting by eq. \ref{fit}.}
\end{figure}

Here we show several dispersion curves obtained in the simulation
and perform their fitting by eq. \ref{fit}. The results are shown
in Fig. \ref{fit-fig}. One can see that eq. \ref{fit} very
accurately describes the dispersion curves obtained in the
simulations. At the same time it efficiently removes the noise.
Because of this we use only the fitted curves in the main text.

\section{Hubbard-Beeby model approximation}

\begin{figure}
\includegraphics[width=8cm]{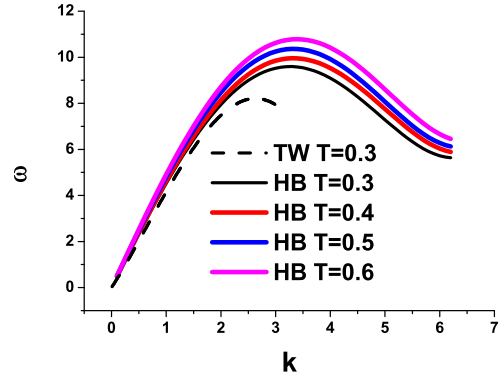}%

\caption{\label{hb} Dispersion curves at $\rho=0.5$ and several temperatures
obtained in Hubburd-Beeby approach. For $T=0.3$ we compare the
results with the results from simulations. From bottom to top:
$T=0.3$ this work (TW), $T=0.3$ HB, $T=0.4$ HB, $T=0.5$ HB,
$T=0.6$ HB.}
\end{figure}

The simplest approach to the excitation spectra in liquids is
Hubbard-Beeby (HB) model \cite{hbmodel}. Although this model does
not give good quantitative agreement with simulation, usually it
gives correct qualitative behavior. In Fig. \ref{hb} we give
several dispersion curves obtained in frames of HB model. We also
give the dispersion curve from this work for $T=0.3$ for
comparison with the HB curve. One can see that the HB model
overestimates the frequencies. But the most important is that the
HB model does not give the decrease of frequencies with
temperature for this system. Therefore HB model is not applicable
to the liquids with anomalous temperature dependence of the
dispersion curves. It is especially interesting because HB model
is widely used to describe the excitation spectra in liquid metals
(see, for instance, Refs. \cite{hb1,hb2,hb3,hb4} and references
therein). As it was shown in Refs. \cite{metpot1,metpot2} the
effective interaction potentials in liquid metals are
core-softened ones as the potential studied in the present work.
Because of this one can expect that HB model gives inappropriate
description of the dispersion curves of liquid metals. However,
nowadays there are more elaborate theories which give much better
description of the dispersion curves of liquids, such as
Generalized Hydrodynamics introduced in Refs. \cite{gh1,gh2} (to
name a few successful implementations see, for instance,
\cite{gh3,gh4,gh5}) and especially the theory which allows to
calculate the dynamic structure factor with high precision
introduced in Refs. \cite{mokshin1,mokshin2,mokshin3} (see, Refs.
\cite{mokshin1,mokshin2,mokshin3,mokshin4,mokshin5,mokshin6} for
some examples of successful implementaltion).

\section{Density anomaly and the speed of sound}

\begin{figure}
\includegraphics[width=8cm]{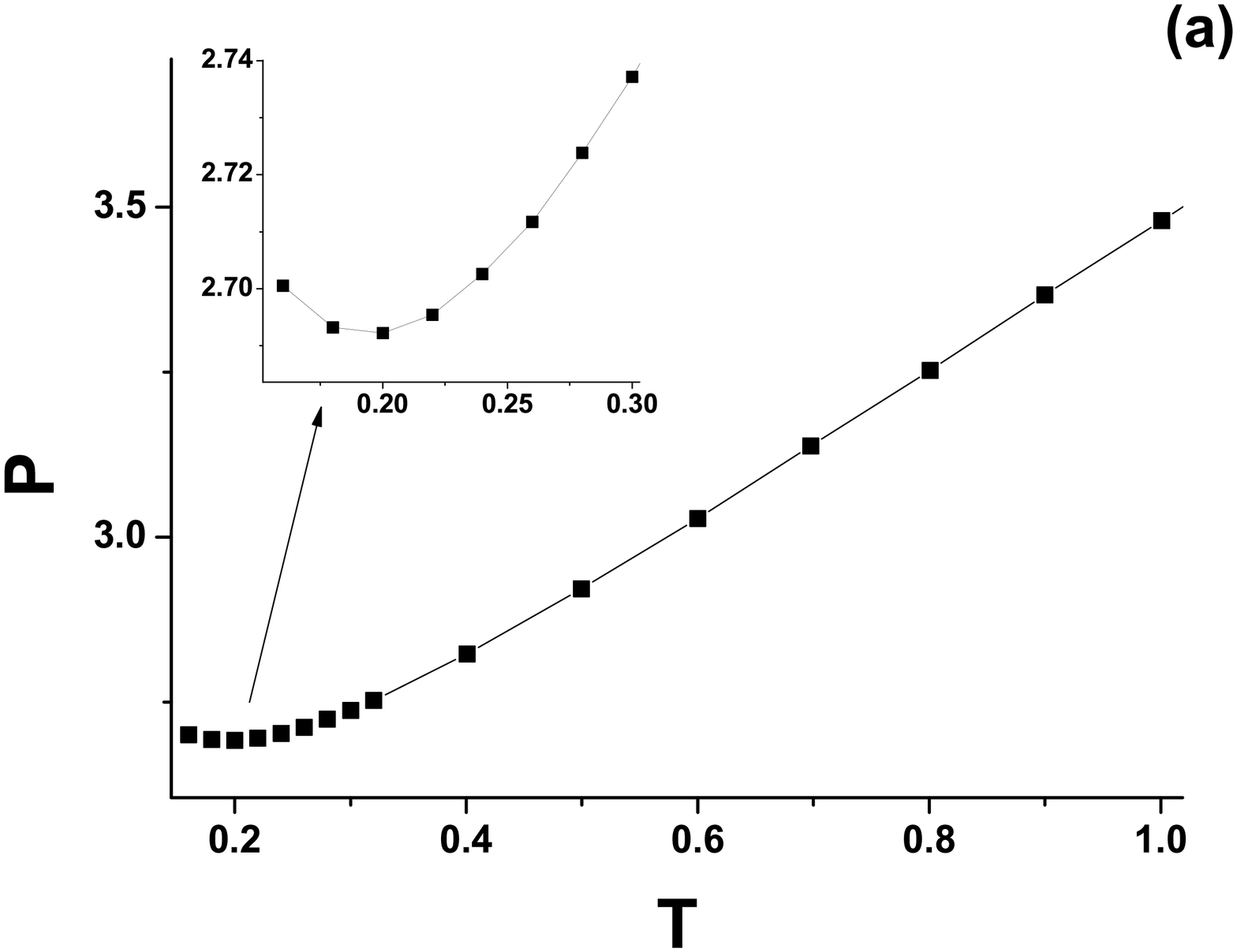}%

\includegraphics[width=8cm]{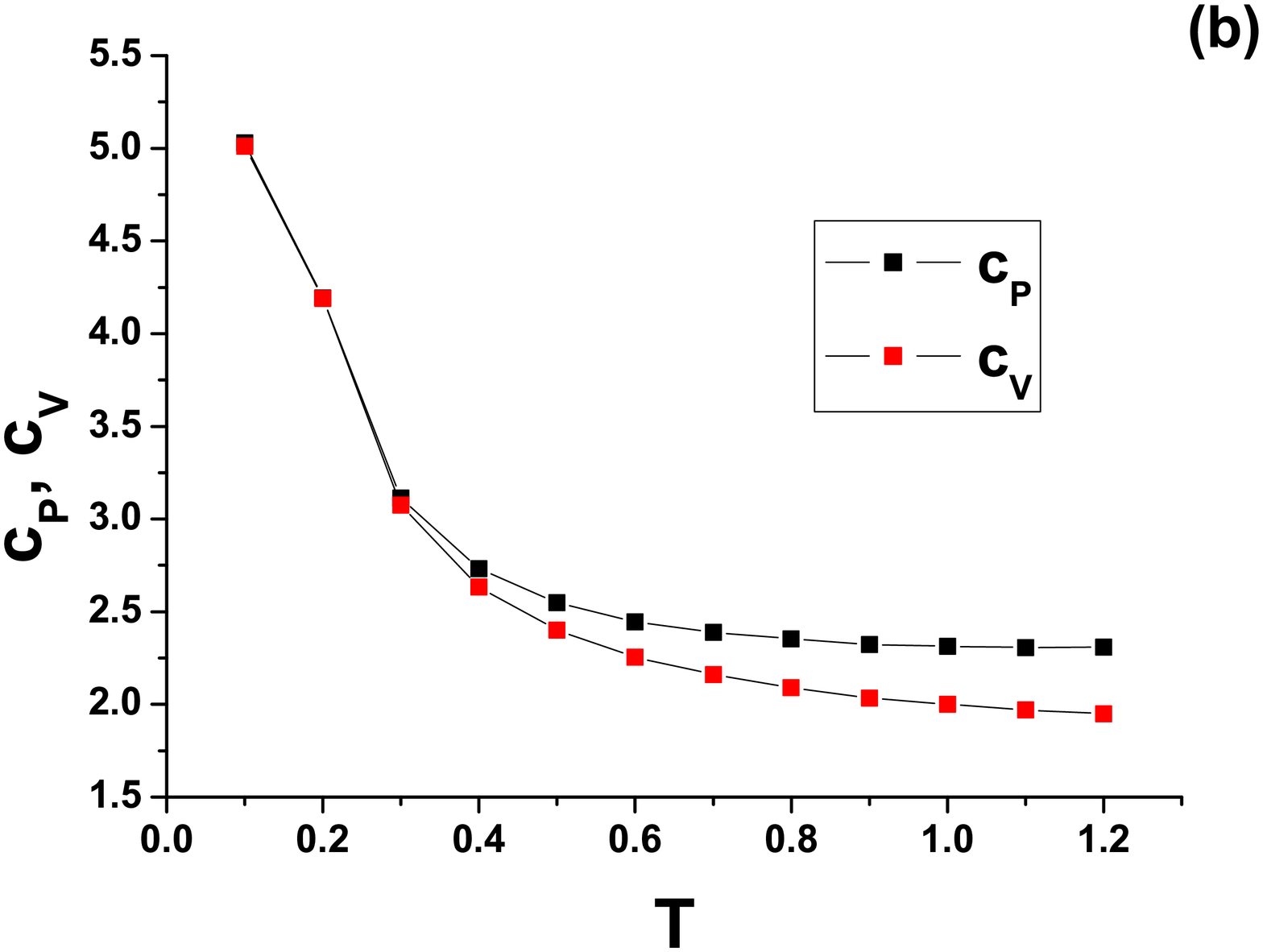}%

\caption{\label{r05-eos}(a) Equation of state along the isochore $\rho=0.5$. The
inset enlarges the region of the density anomaly. (b) Adiabatic
speed of sound at $\rho=0.5$. The symbols are the results of
simulations and the line is the polynomial fit.}
\end{figure}

The SRS system at $\rho=0.5$ the system demonstrates the density
anomaly \cite{s135,s135a,s135b,s135c,s135d,s135e} which leads to
appearance of a minimum on the pressure dependence on temperature
at this isochore. However, the adiabatic speed of sound $c_s$
along the same isochore is a monotonous function of temperature.
The corresponding plots are given in Fig. Fig. \ref{r05-eos} (a)
and (b).

\section{SRS at high density}

\begin{figure}
\includegraphics[width=8cm]{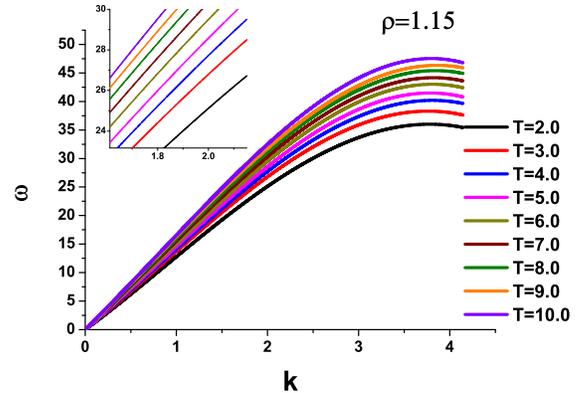}%

\caption{\label{r115}Dispersion curves at density $\rho=1.15$. The inset
enlarges the region about $k \approx 1.9$. One can see that the
frequencies increase with temperature.}
\end{figure}

As a final check of our results we calculate the dispersion curves
of the SRS at high density, where no anomalies is expected. Fig.
\ref{r115} shows the dispersion curves at $\rho=1.15$. One can see
that the dispersion curves behave normally, i.e. the frequency
increases with temperature. Therefore, the anomalous dependence of
the excitation spectra takes place in the region of intermediate
densities where a competition between different length scales
takes place, but it disappears at the high density where the small
length scale is prevailing.

\end{document}